\documentclass[12pt]{emulateapj}
\usepackage{graphicx}

\def\kms{{\rm \thinspace km \thinspace s}^{-1}}

\def\kpc{{\rm\thinspace kpc\ }}

\def\Msun{\hbox{$\rm\thinspace M_{\odot}$}}
\def\pc{{\rm\thinspace pc}}     
       
\def\yr{{\rm\thinspace yr}}

\def\Gyr{{\rm\thinspace Gyr}}

\def\dens{\thinspace\mathrm{g}\thinspace\mathrm{cm}^{-3}}

\def\mnras{MNRAS}
\def\apj{ApJ}
\def\aap{A\&A}
\def\apjl{ApJL}
\def\apjs{ApJS}

\def\araa{ARA\&A}

\shorttitle{Explaining the velocity dispersion of dwarf galaxies by
  baryonic mass loss}
\shortauthors{Matthias Gritschneder, Douglas N.C. Lin}

\begin{document}

\title{Explaining the observed velocity dispersion of dwarf galaxies by
  baryonic mass loss during the first collapse}

\author{Matthias Gritschneder$^{1,2}$,Douglas N.C. Lin$^{1,2}$}
\affil{$^1$ Astronomy and Astrophysics Department, University of California, Santa Cruz,
  CA 95064, USA\\
$^2$Kavli Institute for Astronomy and Astrophysics, Peking
  University,Yi He Yuan Lu 5, Hai Dian, 100871 Beijing, China}
\email{gritschneder@ucolick.org}

\begin{abstract}
In the widely adopted $\Lambda$CDM scenario for galaxy formation, 
dwarf galaxies are the building blocks of larger galaxies.  Since 
they formed at relatively early epochs when the background density 
was relatively high, they are expected to retain their integrity as 
satellite galaxies when they merge to form larger entities. Although
many dwarf spheroidal galaxies (dSphs) are found in the galactic halo
around the Milky Way, their phase space density (or 
velocity dispersion) appears to be significantly smaller than 
that expected for satellite dwarf galaxies in 
the $\Lambda$CDM scenario. In order to account
for this discrepancy, we consider the possibility that they may 
have lost a significant fraction of their baryonic matter content 
during the first infall at the Hubble expansion turnaround.
Such mass loss arises naturally due to the feedback by relatively
massive stars which formed in their centers briefly before the maximum
contraction. Through a series of
N-body simulations, we show that the timely loss of a significant 
fraction of the dSphs initial baryonic matter content can have 
profound effects on their asymptotic half-mass radius, velocity 
dispersion, phase-space density, and the mass fraction between residual 
baryonic and dark matter.  
\end{abstract}


\section{Introduction}

Dwarf galaxies play an important role in the $\Lambda$CDM scenario 
for galaxy formation. They form early, in relatively dense background
environments, and are the building blocks of larger galaxies. 
$\Lambda$CDM simulations show that these dwarf 
galaxies retain their integrity during their merger process and 
predict a rich population of satellite dwarf galaxies around 
large galaxies such as the Milky Way
\citep[e.g.][]{Diemand:2008ys,Springel:2008cr}.

Many satellite dwarf spheroidal galaxies (dSphs) in the 
Galactic halo have been found through SDSS and other surveys.  
Follow-up spectroscopic observations reveal these dSphs are
surrounded by dark-matter halos. But, the internal phase 
space density of the most newly discovered (faint) satellite dSphs 
is significantly smaller than that predicted from the 
dark-matter-only simulations.  This discrepancy poses a 
challenge to the standard $\Lambda$CDM scenario in its 
simplest form
\citep[e.g.][]{Boylan-Kolchin:2011zr,Ferrero:2011dq,Rashkov:2012ly,Boylan-Kolchin:2012bh,Wolf:2012kx}.
Recent observations show that this problem is not limited to the
Milky Way, but present in Andromeda as well
\citep[e.g.][]{Tollerud:2012uq}.

We examine the dSphs' structural adjustment associated with the 
loss of baryonic matter due to star formation feedback.  Today,
the dSphs' total dark to baryonic matter ratio is generally 
much larger than the corresponding cosmological value
\citep{Mateo:1998xi}.
The preferential loss of baryonic matter may be the potential 
solution to both the ``missing baryonic matter'' and
the ``missing satellite'' \citep[e.g][]{Klypin:1999yq,Moore:1999vn} puzzles.

There is very little detectable molecular or atomic gas inside the 
dSphs today.  In \S \ref{basics}, we briefly discuss the possibility that 
a significant fraction of the initial gas content may be lost
due to the feedback effect of first generation stars in them.  
Presumable, this effect is most intense during the collapse 
following the dSph progenitors' turn around from the Hubble flow.
We show, in \S \ref{results}, how the loss of a small fraction of the total
mass may significantly affect dSphs' asymptotic dynamical structure.
Baryonic matter may also be removed from the dSphs through 
tidal or ram pressure stripping by tenuous gas in the 
Galactic halo \citep{Lin:1983ys}. However, gradual loss of gas
from the dSphs' virialized potential would not significantly affect
their internal dynamical structure.

There have been several attempts to include baryonic 
matter in comprehensive cosmological simulations of 
galaxy formation \citep{Gnedin:2009jh}. For computational 
efficiency, a simple-to-use prescription for star formation 
rates (as a function of local gravity, gas density, temperature, 
metallicity, and background radiation) is clearly technically 
desirable. However, there are many competing effects 
including atomic processes, radiation transfer, hydrodynamics and
magnetic fields, which are sensitive to the diverse initial 
and boundary conditions over a large dynamic range in spacial extent 
and time scales. The determining factors for the onset, efficiency, 
and impact of star formation remain enigmatic. 

If the present-day stellar content is distributed 
widely in the form of initial gas through dSphs, the background 
UV flux after the epoch of re-ionization would easily ionize 
the gas within typical dwarf galaxies, delaying or even preventing 
cooling and star formation within them \citep{Dong:2003jb, 
Brown:2012xz}. Most dSphs within the Local 
Group are observed to contain multiple generations of stars, 
the oldest of which formed in the early epochs of cosmic evolution, when 
the background UV flux may have been intense. Possible solutions 
to this conundrum include that the formation of these stars preceded the epoch
of re-ionization or that these dSphs may have lost most (up to $\sim 90 \% $) 
of their original baryonic content which provided a more effective
shield against external UV photons. The second possibility is in line 
with the large mass ratio between dark and baryonic matter in dSphs.  

Most recent cosmological simulations \citep{Brooks:2012zr,Governato:2012ly}
produce results which appear to be in agreement with our baryonic mass loss
conjecture.  It has also been argued that about 10\% of the haloes in 
simulations might not face the 'too big too fail' problem 
\citep{Purcell:2012ve}. However, this conjecture only shifts 
the problem at hand towards the uniqueness of the Milky Way.
Here, we present an idealized toy model for the formation of dSphs. 
We introduce a prescription to link the loss of the interstellar medium
with the burst of first generation star formation in dSphs.  The main 
advantage of our approach here is that it enables us to identify
the key physical effects which determine dSphs' asymptotic dynamical
structure. On the technical side, it also provides adequate
resolution on the scale of the dSphs.  

We first review, in \S \ref{basics}, the basic assumptions, 
methodology, initial and boundary conditions used in our 
numerical models. We then continue
to present the results of our dimensionless simulations (\S
\ref{results}). In \S \ref{discussion} we convert these numbers to
physical quantities and in \S \ref{conclusions} we draw the conclusions.

\section{Basic Approach and Initial Conditions}
\label{basics}

The main cause of the loss of ordinary matter may be feedback associated with
star formation.  In previous simulations \citep{Dong:2003jb} on
the effects of radiative transfer and photoionization in dSphs, it
was shown that star formation can indeed proceed in their more massive and gas rich 
progenitors. The star formation and gas retention efficiency may vary 
widely in galaxies with similar dark matter potentials because they 
depend on many factors, such as the baryonic fraction, external 
perturbation, initial mass function, and background UV intensity. 
The presence of very old stars in dSphs indicates that 
their initial baryonic-to-dark matter content was comparable to the 
cosmic value. This constraint suggests that the initial density 
fluctuation of baryonic matter may be correlated with that of the 
dark matter. For the more massive dwarf elliptical galaxies, 
the star formation efficiency and gas retention rate are much higher. 
Their mass-to-light ratio is regulated by star formation feedback and 
is expected to be nearly independent of their absolute luminosity. The 
results of our theoretical models reproduce the observed
$(M/L)-M_{\rm v}$ correlation \citep{Mateo:1998xi}.

Many dwarf galaxies contain stars with diverse [Fe/H] and some 
contain multiple generations of stars, including 
a recent ($\approx\Gyr$) episode in the Fornax dSph
\citep{de-Boer:2012mq}. 
Significant iron dispersion in dSphs suggests they were self 
contaminated by the retained and recycled supernova ejecta 
from massive stars. But, the total metal content in 
these systems is much less than expected from the heavy-element 
production of massive stars in each episode of star formation. 
Such a deficiency implies that a substantial fraction of the 
dSphs' gravitational potential is generally shallow with a modest 
velocity dispersion. The UV flux from a few massive supernova-
progenitor stars would be adequate to ionize the residual 
gas and evaporate it from the dSphs \citep{Noriega-Crespo:1989ez}.
In the presence of an  initially disturbed gas distribution 
(due to photoionization), only a few percent of the gas enriched 
by the supernovae remains in the center of dSphs \citep{Fragile:2003wm}.

The ionization and supernova blast wave only directly affect
the dynamics of baryonic matter. Although the ejection speed
may substantially exceed the escape speed of the dSphs' dark matter
halo, the loss of baryonic matter makes small fractional changes
in their total mass and potential.  If this mass loss occurs
after the dark matter halo has already virialized, it will only modify the
dark matter content by a small amount. However, we are able to show
here that a timely loss of baryonic matter during the virialization
process would lead to an extended halo structure which is prone to
subsequent harassment. 

During the initial infall of the dSphs, both dark 
and baryonic matter attain infall velocities comparable to their
escape speed before virialization. In the absence of any feedback, the
gravity of the dark matter potential accelerates both
components to attain a kinetic energy which is slightly less 
than their gravitational energy.  As the density of
both dark and baryonic matter increases, gas in the dSphs 
become self shielded against the background UV radiation.
Thermal instability leads to the formation of dense cores
and young stars. Photoionization
and supernovae feedback induces the ejection of 
gas from the central regions of dSphs.  At this advanced
stage of infall, the total energy of the infalling dark 
and baryonic matter is a small fraction of the instantaneous
potential energy \citep{Aarseth:1988lp} and a small fractional
mass loss could offset the near cancellation 
between the kinetic and potential energy. 

We consider the effect of baryonic matter loss on 
the evolution of the dark matter halo in dSphs, especially
during the infall stage of their formation.  In order to illustrate
this effect, we consider a series of N-body simulations
with the publicly available SEREN code \citep{Hubber:2011bu}.
The dark and baryonic matter is represented by 10000 particles 
which only interact with each other through their mutual gravity.
For computational simplicity, we represent the Hubble expansion
turn around with systems of particles at rest with a uniform 
initial density.

The system contracts under its own gravity.  By reducing the mass
of individual particles, we impose an instantaneous loss of a
prescribed fraction $f$ of the total mass
after the half mass radius $R_{\rm hm}$ of the systems has reduced to 
$R_{\rm loss}$ from their original total radius $R_0$ by a factor up to 20 
(see \S\ref{results}). Our goal is to determine the asymptotic structure 
of the systems after virialization as a function of $f$ and $R_{\rm loss}$.
Therefore, we need to generate models with the minimum value of 
$R_{\rm hm} < R_{\rm loss}$.  

The minimum value of $R_{\rm hm}/R_0$ is determined by the initial 
fluctuation spectrum. For systems represented by N particles,
its value, due to the global random noise, is $\sim N^{-1/3}$
\citep{Aarseth:1988lp}.  This contraction factor would be limited
in cosmological simulations with inadequate resolution on the 
scale of dSphs. For the entire range of models in \S\ref{results}, it
is adequate to use $N=10^4$. Although the minimum value of
$R_{\rm hm}/R_0$ may reduce with larger values of $N$, the asymptotic 
structure is affected by the global changes in the gravitational 
potential rather than two body relaxation process (i.e. it 
is determined by the values of $f$ and $R/R_0$ rather than N).

For a fiducial model, we neglect 
any mass loss.  In this case, the system relaxes into a virial
equilibrium after a few times of its initial free fall time 
scale $t_{\rm ff}$.  A small fractional mass loss, especially
during the advanced stage of the infall, can lead to 
much more extended relaxed kinematic structure after violent 
relaxation. We run the model for $2t_{\rm ff}$, which is sufficient to
reach a stable half mass radius in the fiducial case.

\section{Results}
\label{results}

Here, we investigate a number of different scenarios. We vary the
precise time of the mass expulsion, the amount of mass ejected as well
as the region where the mass is expelled. 

\begin{figure}
\begin{center}
{\centering 
\includegraphics[width=8cm]{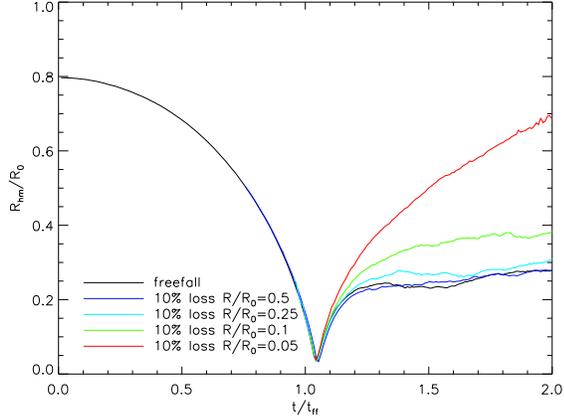}}
\end{center}
\caption{Dependence of the half-mass radius on the precise time of
  loss for different times of loss, characterized by the contraction
  of the halo at the time of the loss. We plot the radius in units of
  the initial radius and the time in units of the free-fall time.
    \label{fig:timebounce}}
\end{figure}

\begin{figure}
\begin{center}
{\centering 
\includegraphics[width=8cm]{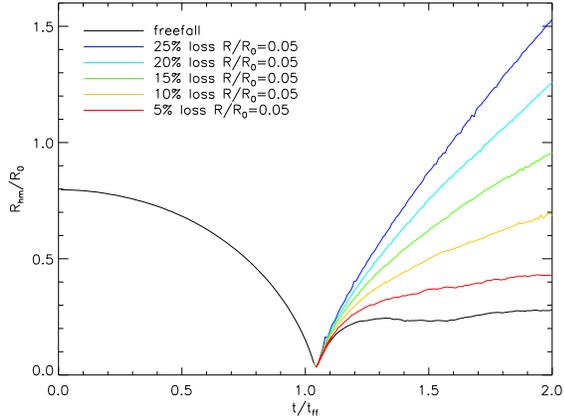}}
\end{center}
\caption{Dependence of the half-mass radius on the amount of loss for
  cases with a loss between 5\% to 25\% of mass loss.  We plot the radius in units of
  the initial radius and the time in units of the free-fall time.
    \label{fig:massbounce}}
\end{figure}

First, we investigate the timing of the loss. We investigate cases
with a contraction to $1/2$, $1/4$, $1/10$ and $1/20$ at the epoch of
the mass loss, respectively. The corresponding dynamical timescales
(to bounce back) are $0.35 t_{\rm ff}$, $0.13 t_{\rm ff}$, $0.03 t_{\rm ff}$, and $0.01 
t_{\rm ff}$.  From Figure
\ref{fig:timebounce} it becomes clear the timing is of crucial
importance. The closer the loss to the maximum contraction, the more
drastic the effect of baryonic mass loss on the entire dark matter
halo. From a physical point of view, the first burst of star formation
and thus the most efficient feedback is indeed to be expected at the
first time high densities are reached, i.e. close to maximum
contraction.

Another interesting feature is the amount of mass loss
needed. Clearly, this parameter is strongly constrained by cosmological
parameters, i.e. the maximum mass loss is limited by the amount of
baryonic matter initially. According to WMAP7 \citep{Komatsu:2011oq},
the ratio of baryonic to dark matter is about 1:4.45. Assuming this
ratio is preserved in the initial halo, there is about 17\% baryons
inside the halo.
We investigate a series of cases with an
loss of 5\% to 25\%. Above 25\% mass loss the halo disperses. Figure
\ref{fig:massbounce} shows a linear dependence of
the final half-mass radius on the amount of mass loss (as expected
from e.g. \citealt{Aarseth:1988lp}). An encouraging result is that
even the loss of about 10\% of the total mass (i.e. slightly more than
half of the baryonic mass) leads to a significant puff up, resulting in an
halo twice a big as in the fiducial case.

In the context of the observed $(M/L)-M_{\rm v}$ relation
  and the final ratio of baryonic to dark matter of about
  1:100 \citep[e.g.][]{Jardel:2012fk}, an even higher fractional 
  mass loss ($\sim 15 \%$) is desirable. However, here we consider 
  only a loss of 10\% during the first infall, which is sufficient 
  to explain the low phase-space density. The observed mass to light 
  ratio is most likely later achieved via star bursts or tidal effects.

\begin{figure}
\begin{center}
{\centering 
\includegraphics[width=8cm]{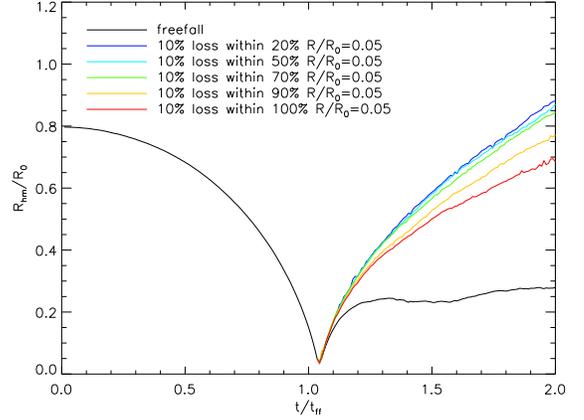}}
\end{center}
\caption{Dependence of the half-mass radius on the region of
  loss. 10\% of the material is lost in to within a range of different
  regions. We plot the radius in units of
  the initial radius and the time in units of the free-fall time.
    \label{fig:centerbounce}}
\end{figure}

\begin{figure}
\begin{center}
{\centering 
\includegraphics[width=8cm]{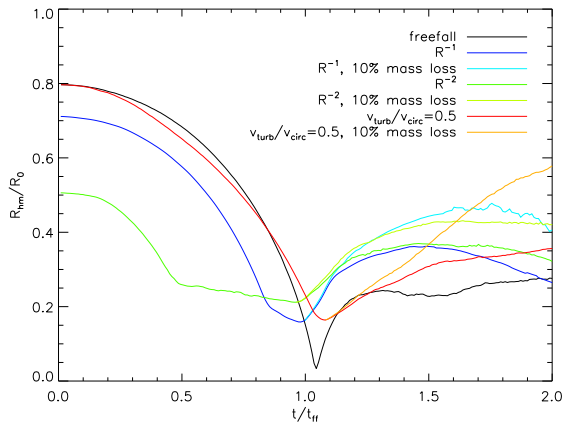}}
\end{center}
\caption{Evolution of the half-mass radius for intial
  conditions with a cusped profile ($R^{-1}$ and $R^{-2}$) or a
  turbulent velocity, respectively.
    \label{fig:cuspvel}}
\end{figure}

\begin{figure}
\begin{center}
{\centering 
\includegraphics[width=8cm]{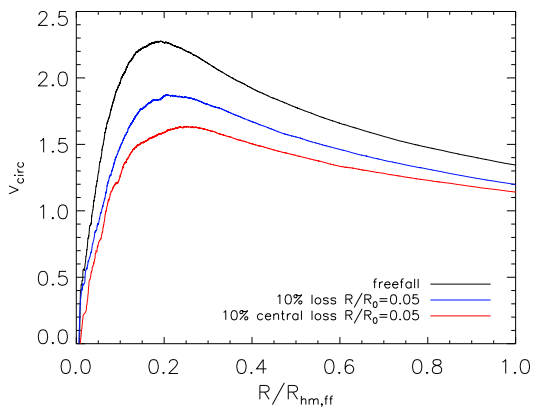}}
\end{center}
 \caption{$v_\mathrm{circ}$, i.e. the potential in three selected
  cases. We plot the velocity in code units versus the radius in
  units of the final half-mass radius in the fiducial, undisturbed
  case.   \label{fig:potential}}
\end{figure}

\begin{figure*}
\begin{center}
{\centering 
\includegraphics[width=16cm]{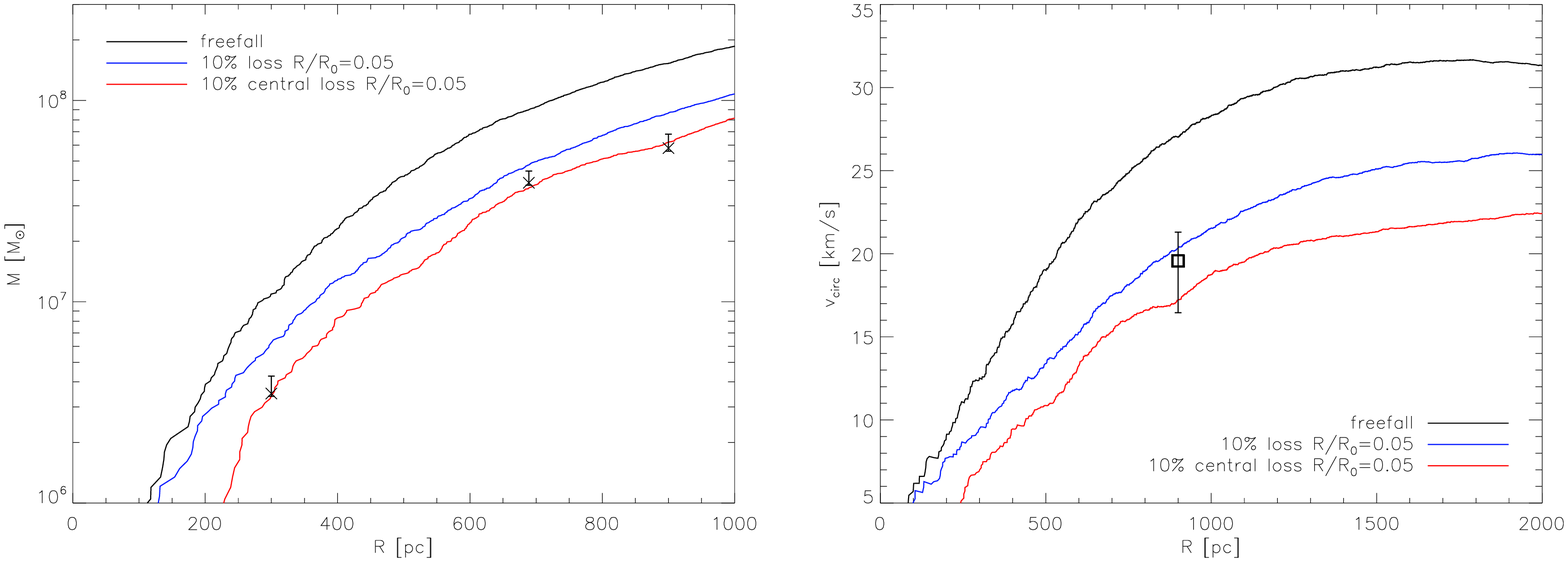}}
\end{center}
\caption{Mapping of the simulations to physical quantities. Left:
  Fitting the mass profile to observations by \citet{Jardel:2012fk}. Right:
  $v_\mathrm{circ}$, i.e. the potential, in three selected cases in
  physical units in comparison with the observed value for Fornax
  \citep[e.g.][]{Jardel:2012fk}.}\label{fig:normalize}
\end{figure*}

To increase the applicability of this numerical experiment to the
problem at hand, it is important to constrain the region where the
mass loss occurs. In a dwarf galaxy, the mass loss is only going to
happen in the dense central region, where the baryons are going to form
stars and subsequently remove the gas via their feedback. We simulate
this effect by reducing mass of individual particles in a central 
fraction of the system.  The results in Figure \ref{fig:centerbounce}
show that a more centralized mass loss leads to a more extended 
asymptotic structure and makes the model more relevant for the formation
of dwarf galaxies. As the dependance on the mass loss region 
is not sensitive, it merely demonstrates that an extended virialized 
system, as observed today, does not require mass loss to take place
throughout the halo.

Recent cosmological simulations \citep[e.g.][]{Diemand:2008fk} show
that dwarf galaxies form from the inside out, as in the classic
secondary infall model \citep{Fillmore:1984uq,Bertschinger:1985kx}. To
test the validity of our model 
in this context we investigate the collapse of a cusped profile. We
set up two profiles, corresponding to $\rho\propto R^{-1}$ and
$\rho\propto R^{-2}$ , respectively. As can be seen in Figure \ref{fig:cuspvel},
this leads to much smaller collapse factor. In fact, this is in better
agreement with cosmological simulations \citep[e.g.][]{Diemand:2007vn},
as the simplicfication of a constant density is not applied
there. However, the half-mass radius still is significantly puffed
up in both cases with a cusped profile, similar to our simplified case.

Another caveat in our initial conditions is the fact that the
particles are at rest initially. To test this simplification we
perform a run with a turbulent velocity. The velocity is set to
correspond to half the circular velocity, i.e. the potential (see
Equation \ref{v_circ} below). Again, the maximum contraction is 
smaller than in the fiducial case (Figure \ref{fig:cuspvel}). Still, in the case with mass loss,
a significant change in the half-mass radius can be seen. We therefore
conclude that the simplifications in our initial conditions do not
affect the results strongly.

In order to assess the change of the velocity dispersion, we evaluate 
the halo potential at a time $t = 2 t_{\rm ff}$, i.e. the circular velocity 
\begin{equation}
v_{\rm circ} = \sqrt{\frac{GM}{R}},
\label{v_circ}
\end{equation}
where $G$ is the gravitational constant, and $M$ is the mass inside
the radius $R$. This approximation is adequate for small $R$ where
the system has already virialized.  

In Figure \ref{fig:potential}, we plot $v_{\rm circ}$ as a 
function of $R/R_{\rm hm, ff}$ where $R_{\rm hm, ff}$ is the 
final half mass radius of the fiducial model (represented by
the top black line).  We also compare the fiducial case and 
two models with a 10\% mass loss where the mass loss is 
either confined to the central 20\% (the bottom red line) or 
extended throughout the the entire halo (the middle blue line). 
These results show that mass loss significantly
reduces the distribution of $v_{\rm circ}$ at all radii 
to about $3/4$ of the case without mass loss.

Finally, the halos here all show a cored density profile, as
has already been deduced by \citet{Aarseth:1988lp}. This result
is in agreement with the finding of most recent observational models
which generally favor a cored rather than a cusped profile
\citep[e.g.][]{Amorisco:2012kx}

\section{Discussion}
\label{discussion}

The above models show, that a loss of 10\% of the total mass,
from the central 20\% of the total halo, at a time when the halo has
contracted to $1/20$ of its original size, is sufficient to 
reduce the velocity dispersion by $\sim 20-30 \%$ and to cause 
the halo to attain an extended asymptotic structure.

In order to apply these simulated models in the context of
Galactic dSphs, it is necessary to convert some of the 
dimensionless quantities into physical numbers. A straightforward 
quantity to do the mapping is the mass. By fitting the simulated
and observed mass and velocity dispersion distribution, we can 
establish a characteristic mass, length scaling, and a complete 
normalization set.

We apply this approach to fit the observed mass profile of the 
Fornax dSph. We adopt the model parameters inferred from the
observed velocity dispersion by \citet{Jardel:2012fk}.  They 
determined its mass to be $M_{300}=3.5^{+0.77}_{-0.11}\times10^6\Msun$
inside $R=300\pc$, $M_{689}=3.9^{+0.46}_{-0.11}\times10^7\Msun$ 
inside the projected half-light radius ($R=689\pc$) and
$M_{900}=5.8^{+1.0}_{-0.2}\times10^7\Msun$ inside the 
unprojected half-light radius ($R=900\pc$).  They also obtained
an observed luminosity weighted line-of-sight velocity dispersion
$\sigma=11.3^{+1.0}_{-1.8}\kms$ 

We set the initial mass of Fornax dSph's entire halo to be 
$1.5\times10^9\Msun$ and choose a length scale which matches 
the observationally inferred mass distribution (Figure 
\ref{fig:normalize} left panel). Then, we adopt their value of 
$\sigma$ and convert it to a circular velocity $v_{\rm circ}
=\sqrt{3\sigma^2}$ \citep{Wolf:2010uq} and plot that value at 
the position of the unprojected half-light radius (Figure 
\ref{fig:normalize} right panel).

Although our choice of the initial total mass is somewhat arbitrary, 
the combined constraint of the observed mass at various physical radii 
limits the degree of freedom.  There are several combinations of 
mass and radius normalization factors which can fit the data points 
in the left panel, but they all leave the match between the 
observations and our profiles in the right panel invariant. 
Therefore, we do the conversion with an reasonable choice of $M_{\rm tot}$.

Our choice of normalization also gives a timescale, 
$t_{\rm ff} = 2.2\Gyr$, which puts our simulations in a 
reasonable cosmological context. The length scale is 
$R_{\rm hm,ff}=9.2\kpc$ and the dark matter density is
$\rho_{DM} \sim 3 \times 10^{-26}\dens$.  If we 
use a cosmological baryonic to dark matter ratio, the 
initial mass of baryonic mass would be $\sim 2.55 \times 
10^8 M_\odot$. Despite the loss of more than 50\% of the initial
baryonic mass (which corresponds to a loss of $\sim 10 \%$ of  
the initial total mass), Fornax would still be able to 
have retained a sufficient amount of baryonic matter 
to account for its observed stellar mass
\citep{Mateo:1998xi,Jardel:2012fk}
and some loss of baryonic matter during the subsequent 
stellar and dynamical evolution.

Our best-fit model requires the mass loss to take place 
after the half mass radius has contracted by a factor of 
20.  This collapse factor would reduce the instantaneous
infall time scale to $\sim 10^7\yr$ and the average 
density of the baryonic matter to $\sim 10^{-22}\dens$.
These conditions are likely to lead to the onset of star bursts and
consequently a strong feedback effect. A detailed investigation will
be performed in future work.

\section{Summary}
\label{conclusions}

With the help of a simplified N-body model we are able to show a
possible physical cause for the observed puff-up of satellite 
dSphs around the Milky Way. We propose that the internal star 
formation feedback during advanced stages of the first infall
is sufficient to modify the dark matter potential in order to 
explain their modified phase-space density. 

A natural next step is to use an SPH scheme to carry out a set 
of simulations which include a realistic treatment of the loss
of baryonic matter.  With the SEREN code, we can introduce
a more realistic prescription for radiative or supernova driven loss
of baryonic matter.  For example, the impulsive mass loss prescription
may be relaxed with a model for star formation which proceeds on a time
scale comparable to the dynamical time scale during the advanced 
stage of infall. Two types of particles may be used
to represent gas and dark matter.  Star formation may mostly
occur in the baryonic matter concentrated regions. We 
plan to include the Galactic potential which removes
dark matter venturing beyond the dSphs tidal radius.  
Recent simulations \citep[e.g.][]{Pawlik:2012qf} include some of these
effects and show that the potential is indeed reduced in simulations
with feedback \citep{Brooks:2012zr}. 

Here, we focus on the physical effect. During the first infall, the
particles (dark matter and gas) are on near parabolic orbits. With the
removal of a fraction of the mass, preferentially in the centre, the
potential is changed sufficiently to make some particles orbits
become unbound. We are able to show that e.g. for a loss of 10\% of the
material in the central 20\% of the (current) halo the observed lower
value of the potential/the velocity dispersion can be explained within the
framework of $\Lambda$CDM.  The asymptotic dark matter 
distribution in these dSphs generally has a cored profile.

\section{Acknowledgements}

We thank D. Hubber for providing and explaining the SEREN code and
J. Bullock for useful conversation. This work was supported in part by NASA
grant NNX08AL41G. 
M.G. acknowledges funding by the Alexander von Humboldt Foundation 
in the form of a Feodor-Lynen Fellowship and additional funding by the China
National Postdoc Fund Grant No. 20100470108 and the National Science
Foundation of China Grant No. 11003001 during the initial stages of
this project.

\bibliographystyle{apj}

\end{document}